# RED behavior with different packet sizes

Stefaan De Cnodder, Omar Elloumi[*], Kenny Pauwels

*Traffic and Routing Technologies project*
*Alcatel Corporate Research Center, Francis Wellesplein, 1 - 2018 Antwerp, Belgium*

*Abstract*

*We consider the adaptation of random early detection (RED) as a buffer management algorithm for TCP traffic in Internet gateways where different maximum transfer units (MTUs) are used. We studied the two RED variants described in [4] and point out a weakness in both. The first variant where the drop probability is independent from the packet size discriminates connections with smaller MTUs. The second variant results in a very high packet loss ratio (PLR), and as a consequence low goodput, for connections with higher MTUs. We show that fairness in terms of loss and goodput can be supplied through an appropriate setting of the RED algorithm.*

*Index Terms:* RED, TCP, active queue management.

## 1/ Introduction

The random early detection (RED) algorithm is becoming a *de-facto* standard for congestion avoidance in the Internet and other packet switched networks. RED is an active queue management algorithm that aims at increasing the overall network throughput while maintaining low delays. The main idea behind RED is to prevent from packets being dropped because of buffer overflow by dropping them randomly when the average queue size is above a certain threshold. When packets are dropped from tail –the default drop strategy in packet switched networks– losses are arbitrarily distributed among different competing flows. By randomly dropping packets RED aims at fairly distributing losses in proportion to the amount of bandwidth used by each flow. Another goal of RED is to avoid global synchronization caused by dropping packets in bursts, as it is the case for tail-drop gateways. Adaptive sources (e.g. TCP) reduce their sending rate as a reaction to packet drops, which are considered as an implicit indication of congestion. A more efficient way, referred to as early congestion notification (ECN), to inform sources of the congestion is to mark packets rather than dropping them [9]. This explicit way aims at preventing the extra delay incurred by packet retransmission.

As a consequence of the incremental deployment of RED, several algorithms based on RED have been and are still being proposed to improve its performance (e.g. [3,6]). The authors of [2] proposed RED with In and Out (RIO) as an extension to RED to discriminate low priority packets (Out) in times of congestion. It is expected that a variant of RIO will be used in differentiated services (DiffServ) networks as a means to provide different forwarding treatments for different packet priorities.

RFC2309 [1] states that RED should be used as the default mechanism for managing queues in routers unless there are good reasons to use another mechanism. To this end, strong recommendations for testing, standardization and widespread deployment of active queue management in routers, to improve the performance of today's Internet are made.

---

[*] Corresponding author: email: omar.elloumi@alcatel.be, Tel: +32 3 240 78 33, Fax: +32 3 240 99 32

In this paper we study the two RED variants proposed in [4] and point out a weakness in both. The first variant where the drop probability is independent from the packet size discriminates connections with smaller MTUs. The second variant results in a very high Packet Loss Ratio (PLR), and as a consequence low goodput, for connections with higher MTUs. We propose 3 other variants of RED and compare their performance to the variants proposed in [4]. We show that the proposed variants solve the weaknesses pointed out in the original RED variants.

This paper is structured as follows: Section 2 describes the main idea behind RED as well as the two proposed RED variants described in [4]. Section 3 reports simulation results of RED when the traffic is generated by TCP connections with different MTU values. Section 4 describes the proposed modifications in order to improve the performance of RED. Section 5 reports simulation results showing the obtained goodput and PLR for the proposed RED variants. Finally, Section 5 gives conclusions and recommendations on the optimal configuration of RED.

## 2/ RED and TCP background information

In this section we describe the RED algorithms as well as the two variants proposed in [4]. Then we briefly introduce the TCP congestion control mechanisms. The interested reader can find additional details on TCP congestion control in [10].

### 2.1/ RED gateways

In order to allow transient bursts, RED randomly drops packets based on the average queue rather than on the actual one. The average queue size is estimated as follows: $avg \leftarrow (1-w_q) \cdot avg + w_q \cdot q$, where $avg$ is the average queue size, $w_q$ is the weight used for the exponential weighted moving average (EWMA) filter and $q$ is the instantaneous queue size. An arriving packet is unconditionally accepted if the average queue size ($avg$) is below a minimum threshold, is dropped with a certain probability if $avg$ is between the minimum and a maximum threshold, and finally dropped otherwise. In [4] two variants of RED are proposed, the first one (that we denote by RED_1) does not take the packet size into account when estimating the drop probability, while the second (that we denote by RED_2) weights the drop probability by the packet size. This kind of discrimination between small and large packets is intended to avoid extra delay, incurred by retransmissions, for delay sensitive interactive traffic (e.g. Telnet) which generally consists of small packets. Table 1 gives the steps needed in order to estimate the drop probability, $p_a$, on each packet arrival for RED_1 and RED_2.

| RED_1 | RED_2 | Step |
|---|---|---|
| $count \leftarrow count + 1$ | $count \leftarrow count + 1$ | (1) |
| $p_b \leftarrow \max p \cdot \dfrac{avg - \min_{th}}{\max_{th} - \min_{th}}$ | $p_b \leftarrow \max p \cdot \dfrac{avg - \min_{th}}{\max_{th} - \min_{th}}$ | (2) |
| - | $p_b \leftarrow p_b \cdot L/M$ | (3) |
| $p_a \leftarrow p_b/(1 - count \cdot p_b)$ | $p_a \leftarrow p_b/(1 - count \cdot p_b)$ | (4) |

Table 1: necessary steps to compute the drop probability, $p_a$, for RED_1 and RED_2

In Table 1 the significance of the used parameters and variables is as follows: $p_b$ is a temporarily dropping probability, $\max_p$ is an upper bound on the temporarily packet drop probability, $\min_{th}$ and $\max_{th}$ are the two thresholds limiting the region where packets are randomly dropped, $L$ is the size of the incoming packet, $M$ is the maximum packet size and *count* is the number of accepted packet since the last drop or since *avg* exceeded $\min_{th}$. Note that the only difference between the two algorithms is the third step in RED_2 where the temporarily dropping probability $p_b$ is weighted by the packet size.

Step (2) aims at having a drop probability that increases linearly, from 0 to $\max_p$, as the average queue size increases.

An attractive property of RED_1 resulting from using the *count* variable is that the number of accepted packets between two packet drops is uniformly distributed [4]. By having a uniform distribution, packet drops are not clustered, avoiding again possible synchronization of TCP sources. Although quantitative benefits of having a uniform distribution were not, at the best of our knowledge, reported in the literature it is commonly admitted that having light-tailed distributions (such as the uniform distribution) gives better performance in terms of efficiency and fairness[1]. The proof given in [4] is as follows: let X be the number of packets that arrive after a dropped packet and until the next packet is dropped then:

$$P[X = n] = \frac{p_b}{1 - (n-1) \cdot p_b} \cdot \prod_{i=0}^{n-2}\left(1 - \frac{p_b}{1 - i \cdot P_b}\right) \quad (1)$$
$$= p_b \text{ for } 1 \leq n \leq 1/p_b$$
$$\text{and } P[X = n] = 0 \text{ for } n > 1/p_b.$$

Note that the interval between two drops is not uniformly distributed for RED_2. In the following section we propose a modification to RED_2 in order to solve this problem.

**2.2/ TCP congestion control mechanisms**

The slow start algorithm [10] was proposed by Jacobson as a congestion avoidance and control algorithm for TCP after a congestion collapse of the Internet. This algorithm introduces a congestion window mechanism to control the number of bytes that the sender is able to transmit before waiting for an acknowledgment. For each received acknowledgment, two new segments may be sent. When the window size reaches a threshold value, SSThreshold, the algorithm operates in Congestion Avoidance mode. The slow start is triggered every retransmission timeout by setting SSThreshold to half the current congestion window and the congestion window to one segment. In the Congestion Avoidance phase, the congestion window is increased by one segment every Round Trip Time (RTT). Thus when the mechanism anticipates a congestion, it increases the congestion window linearly rather than exponentially. The upper limit for this region is the value of the receiver's advertised window. If the transmitter receives three duplicate acknowledgments, SSThreshold is set to half the preceding congestion window size while this latter is set to one packet for TCP Tahoe and half the previous congestion window for TCP Reno. At this point the algorithm assumes that a packet is lost, and retransmits it before the timer expires. This algorithm is known as the *fast retransmit fast recovery* mechanism and avoids the inactivity period before the expiration of the retransmission timer.

---

[1] Unpublished simulation results performed by Sally Floyd show that light-tailed distributions give better performance than heavy-tailed ones. See the note in http://www.aciri.org/floyd/REDdistributions.txt for more details.

# 3/ Simulations with different packet sizes

In this section we show simulation results obtained when the traffic is generated by TCP sources with different packet sizes. Our simulations are performed using the two variants of the RED algorithm described in 2.1.

## 3.1/ Simulation model

Our simulation model is composed of 3 groups of TCP sources/destinations sharing the same network path composed of a bottleneck link of 30 Mbits/s connecting two routers. Each group is composed of 20 TCP sources/destinations. All the simulations were done using TCP Reno [10] and TCP with selective acknowledgments [9]. The MTU for each group is respectively 1500, 750 and 375 bytes[2]. The timeout granularity was set to 200 ms. We performed two sets of simulations with small and large propagation delay values for the bottleneck links (15 ms and 80 ms). The simulation model is depicted in Figure 1.

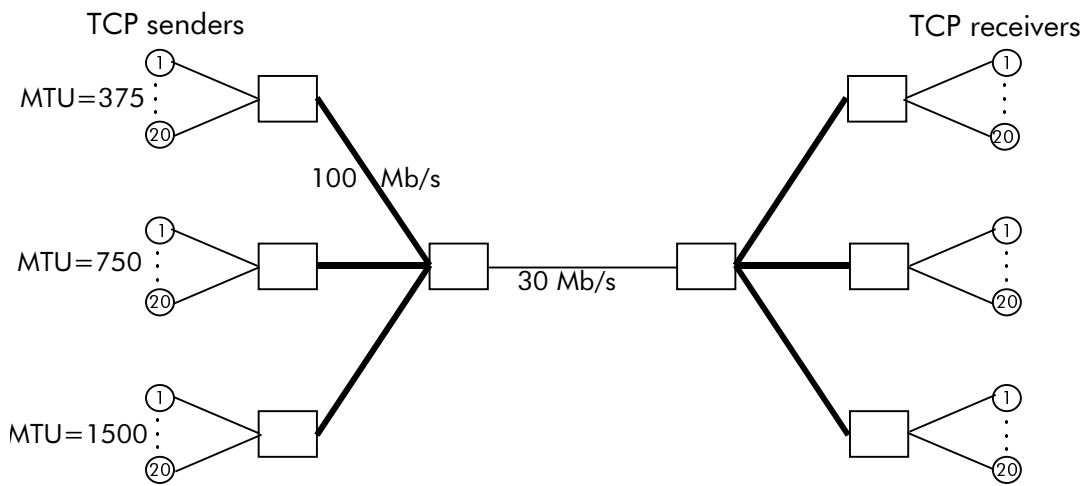

Figure 1: The simulation model

## 3.1/ Simulation results

Simulation results reporting the goodput and the PLR for two values of the propagation delay are depicted in the following figures. For goodput results we plot the sum of the goodput obtained by each of the 20 TCP connections having the same MTU (see Figure 2 and Figure 4). A PLR for each MTU value is reported in Figure 3 and Figure 5. The PLR is defined as the number of dropped packet with a given MTU over the total number of packets having that MTU.

From the simulation results we can conclude that RED_1, which drops packets without taking into account their size, results in a higher throughput for large packets. The obtained goodput is consistent with the TCP goodput estimation formula [8] (see Section 4, equation (3)): the goodput doubles when the MTU is doubled. However small Telnet packets could have a too high a PLR from an interactive application requirements point of view.

---

[2] Although these MTU values may be uncommon we choose them to better explain the results of the paper. These MTU values correspond to large, medium and small MTUs in the graphs.

For RED_2 we can clearly see that the PLR is very large for large MTU values. This leads to a considerable degradation of the TCP goodput especially when the propagation delay is small (Figure 2). Due to an increased PLR for high MTU values (close to 14% for large MTUs), the number of timeouts increases considerably and leads to important intervals of inactivity waiting for a timeout to expire. The high value of the PLR for large MTUs prevents the congestion window from reaching a sufficient value to trigger the *fast retransmit fast recovery* algorithm after a packet loss. This prevents connections with a high MTU values from achieving a considerable amount of goodput. This problem is less important when the propagation delay is relatively high (see Figure 4).

We can see from Figure 3 and Figure 5 that in all cases TCP SACK experiences a higher PLR compared to TCP Reno. For RED_2 the goodput collapse observed for large packet sizes is more severe for TCP SACK which normally gives better performance compared to TCP Reno (see Figure 2). This is because TCP SACK is more aggressive during the recovery phase after packet losses. However, comparing the performance of TCP SACK and TCP Reno is out of the scope of this paper. Our intention is to show that the RED behavior with the presence of different packet sizes is not specific for a given TCP version, i.e. the results obtained for both versions are comparable.

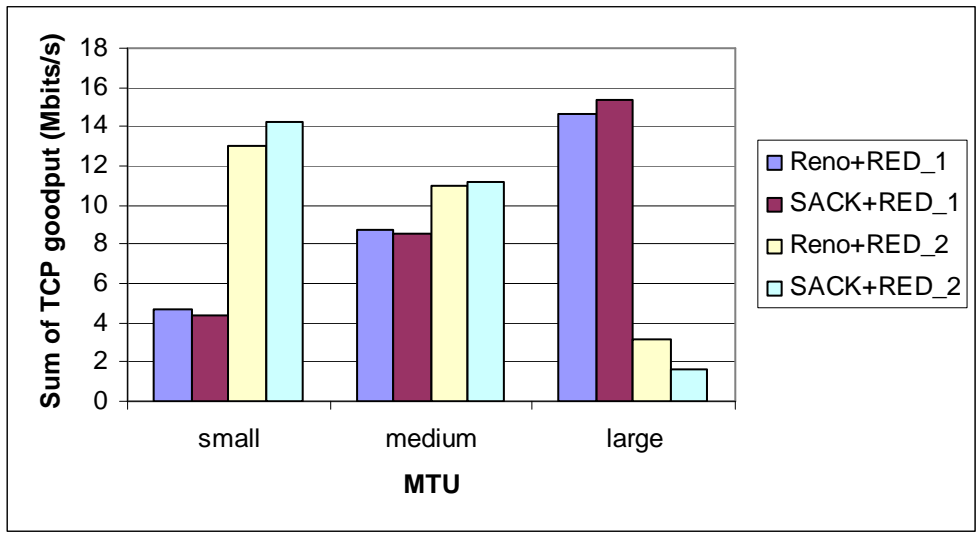

Figure 2: Obtained goodput for different MTUs, small propagation delay

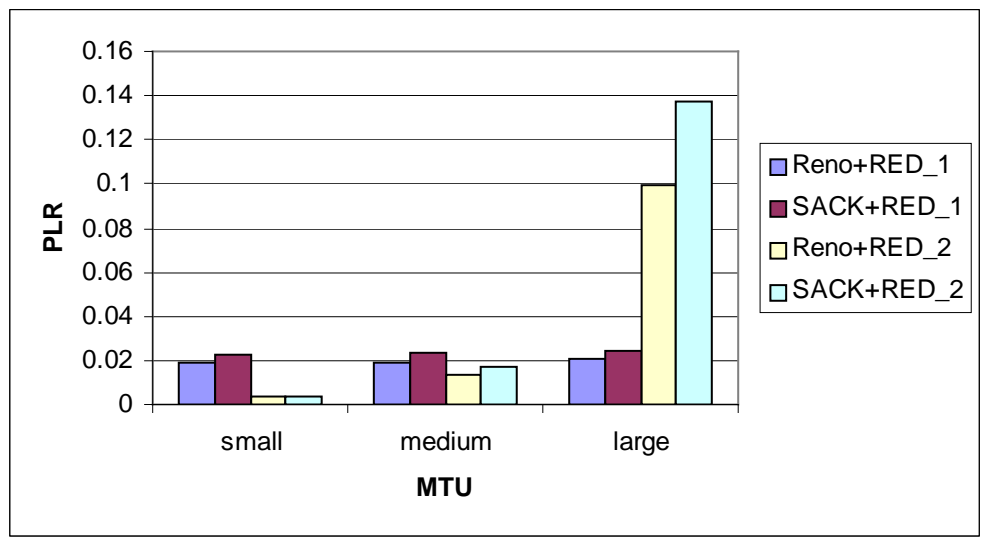

Figure 3: PLR for different MTUs, small propagation delay

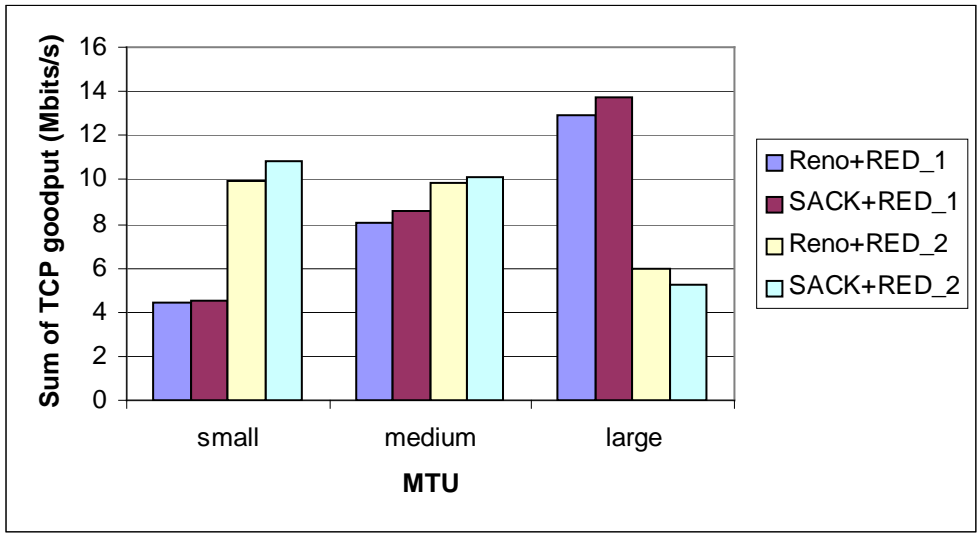

Figure 4: Obtained goodput for different MTUs, large propagation delay

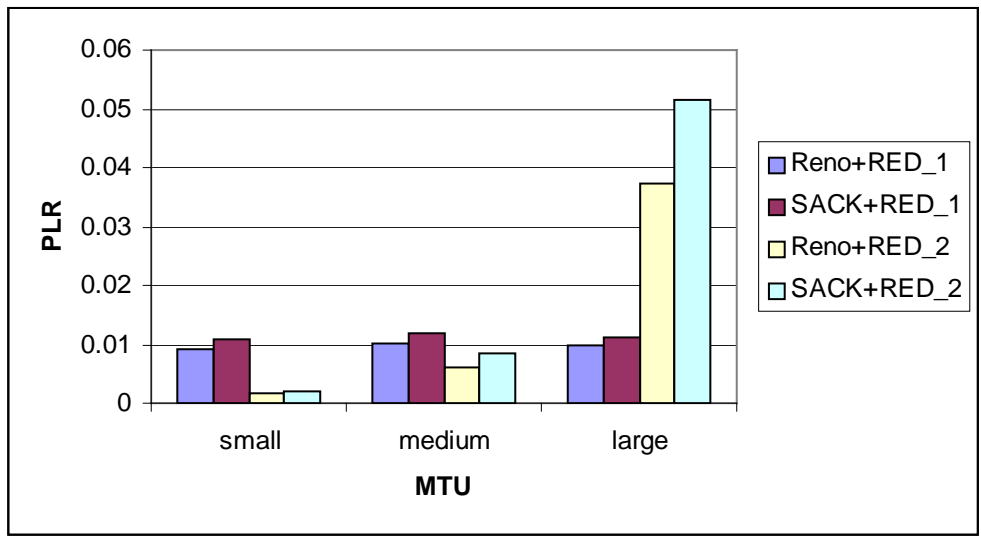

Figure 5: PLR for different MTUs, small propagation delay

## 3/ Proposed modifications to RED

In this section we propose three new settings of the RED algorithm: RED_3, RED_4 and RED_5. The main differences compared to RED_1 and RED_2 is the way in which the drop probability is estimated. The following table explains the basic steps for calculating the drop probability for the three new proposed variants.

| RED_3 | RED_4 | RED_5 | Step |
|---|---|---|---|
| $count \leftarrow count + 1$ | - | - | (1) |
| $p_b \leftarrow \max p \cdot \dfrac{avg - \min_{th}}{\max_{th} - \min_{th}}$ | $p_b \leftarrow \max p \cdot \dfrac{avg - \min_{th}}{\max_{th} - \min_{th}}$ | $p_b \leftarrow \max p \cdot \dfrac{avg - \min_{th}}{\max_{th} - \min_{th}}$ | (2) |
| - | - | - | (3) |
| $p_a \leftarrow \dfrac{p_b \cdot L}{(1 - count \cdot p_b) \cdot M}$ | $p_a \leftarrow \dfrac{p_b \cdot L}{(1 - count \cdot p_b) \cdot M}$ | $p_a \leftarrow \dfrac{p_b}{(1 - count \cdot p_b)} \cdot (\dfrac{L}{M})^2$ | (4) |
| - | $count \leftarrow count + \dfrac{L}{M}$ | $count \leftarrow count + (\dfrac{L}{M})^2$ | (5) |

Table 2: necessary steps to compute the drop probability, $p_a$, for RED_3, RED_4 and RED_5

RED_3 is proposed as an adjustment to RED_2 in order to weight the final packet drop probability by the packet size. The only modification compared to RED_2 is that step (3) is removed and step (4) is modified in order to weight the final drop probability, $p_a$, (rather than the temporary one, $p_b$) by the packet size.

RED_4 is a small modification to RED_3 aiming at conserving a uniformly dropping function by incrementing *count* by $\dfrac{L}{M}$ and moving the update of *count* after the final drop probability calculation.

In order to proof that the number of accepted packets between two drops is uniformly distributed, let $N$ be the number of incoming packets after a packet is dropped until the next drop including this dropped packet and $L_i$, the length of the i[th] incoming packet after a drop then:

$$P[N = n] = \dfrac{p_b \cdot \dfrac{L_n}{M}}{1 - \dfrac{\sum_{i=1}^{n-1} L_i}{M} \cdot p_b} \cdot \prod_{i=1}^{n-1}\left(1 - \dfrac{\dfrac{L_i}{M} \cdot p_b}{1 - \dfrac{\sum_{j=1}^{i-1} L_j}{M} \cdot P_b}\right) \quad (2)$$

$$= p_b \cdot \dfrac{L_n}{M} \text{ if n verifies the following condition: } \sum_{i=1}^{n} \dfrac{L_i}{M} \leq 1/p_b,$$

and $P[N = n] = 0$ if $\sum_{i=1}^{n} \dfrac{L_i}{M} > 1/p_b$.

The reason for which we proposed RED_5 comes from the TCP goodput estimation formula proposed in [8]:

$$goodput \leq \frac{MSS \cdot C}{RTT \cdot \sqrt{p}},  \quad (3)$$

where $C$ is a constant, $MSS$ is the Maximum Segment Size, $RTT$ is the Round Trip Time and $p$ is the packet drop probability. Let $MSS_1$ and $MSS_2$ be two different MSS values corresponding to two TCP connections with the same RTT then in order to achieve fairness the following equation needs to be satisfied: $\frac{p_1}{MSS_1^2} = \frac{p_2}{MSS_2^2}$, where $p_1$ and $p_2$ are respectively the drop probability for the first and the second connection. Substituting $MSS_1$ by the packet size, $L$, and $MSS_2$ by the maximum packet size, $M$, explains step (4) in Table 2.

Note that as RED_4, RED_5 retains the property of a uniform dropping function. The proof is as in equation (1) and results in the following expression of the dropping distribution:

$$P[N=n] = p_b \cdot (\frac{L_n}{M})^2 \text{ if } n \text{ verifies the following condition: } \sum_{i=1}^{n}(\frac{L_i}{M})^2 \leq 1/p_b, \quad (4)$$

$$\text{and } P[N=n]=0 \text{ if } \sum_{i=1}^{n}(\frac{L_i}{M})^2 > 1/p_b.$$

## 4/ Simulation results of the proposed RED variants

Simulation results reporting the goodput and the PLR for the 3 proposed RED variants are depicted in Figure 6 through Figure 13.

We can conclude that RED_3 and RED_4 result in comparable goodput and PLRs and provide a relatively good fairness when the propagation delay is small. This fairness is less acceptable when the propagation delay is large. However in order to improve the throughput larger packets has to be chosen. This means that the throughput increases as a function of the packet size which is a desirable property in order to keep the packet overhead low. The PLR for RED_3 and RED_4 doubles when the MTU doubles.

Finally RED_5 results in a good fairness especially when the propagation delay is large. The PLR is proportional to the square of the MTU which is an expected result. From a theoretical point of view the drop probability should be weighted by the square of the ratio of the packet size over the maximum packet size. The TCP goodput estimation formula given by equation (3) holds under the assumption that all retransmissions are made upon the receipt of three duplicate acknowledgments and not after a timeout. Hence using a small value of the timer granularity and RED_5 dramatically improves the fairness[3] (compared to RED_1) when the traffic is generated by TCP sources having different MTUs.

---

[3] Perfect fairness can hardly be achieved since it is likely that different TCP connections have different RTT values. The fairness, as a function of the RTT, could be improved if the router has a means to estimate the RTT of each connection which, in practice, is hardly feasible.

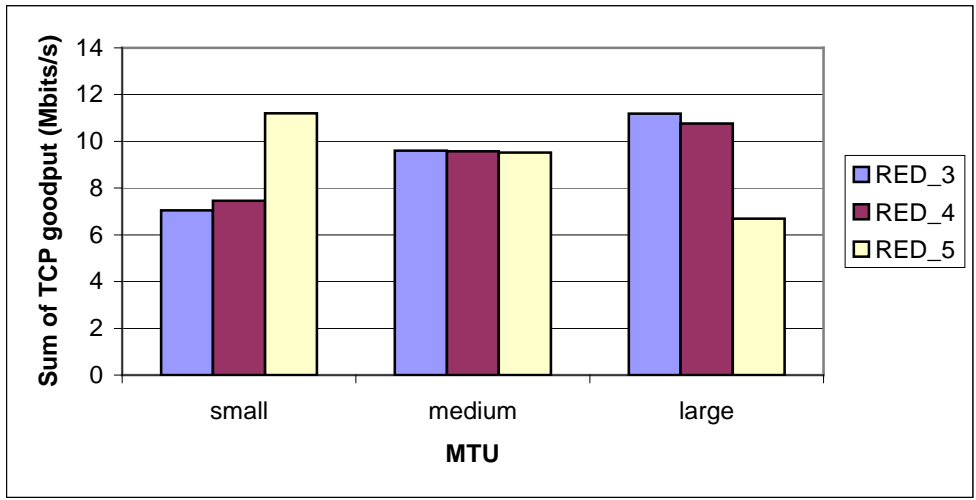

Figure 6: Obtained goodput for different MTUs, small propagation delay, TCP Reno

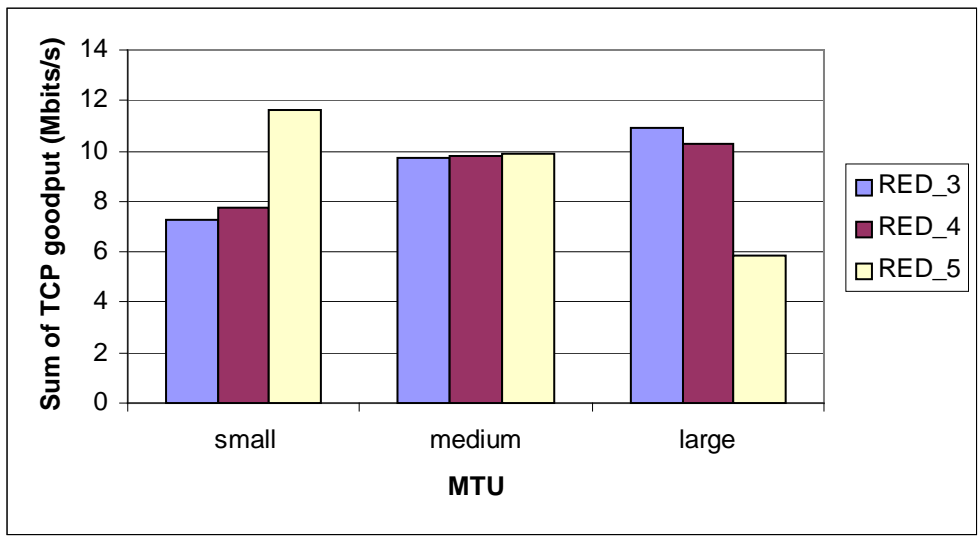

Figure 7: Obtained goodput for different MTUs, small propagation delay, TCP SACK

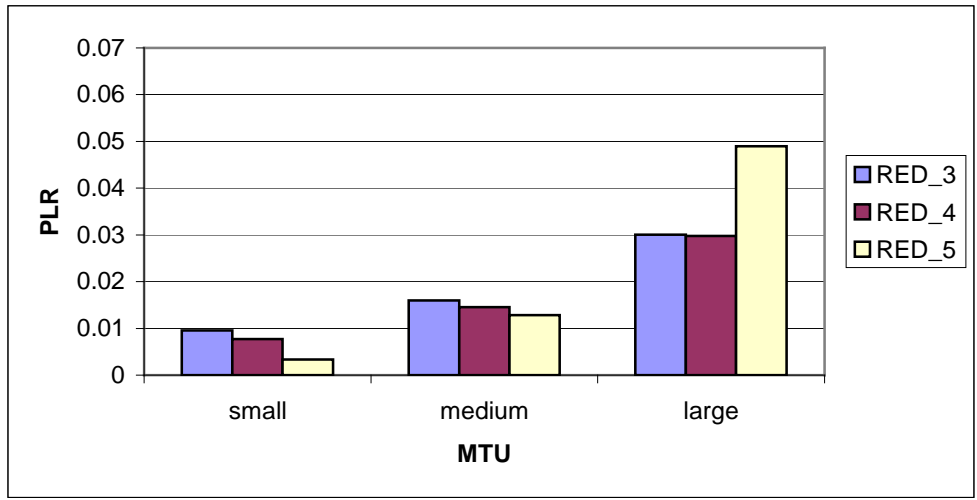

Figure 8: PLR for different MTUs, small propagation delay, TCP Reno

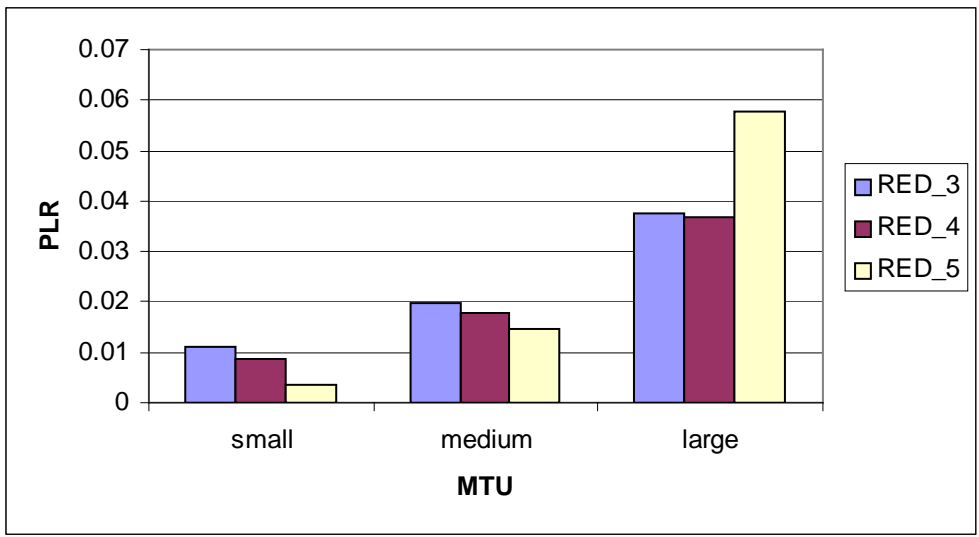

Figure 9: PLR for different MTUs, small propagation delay, TCP SACK

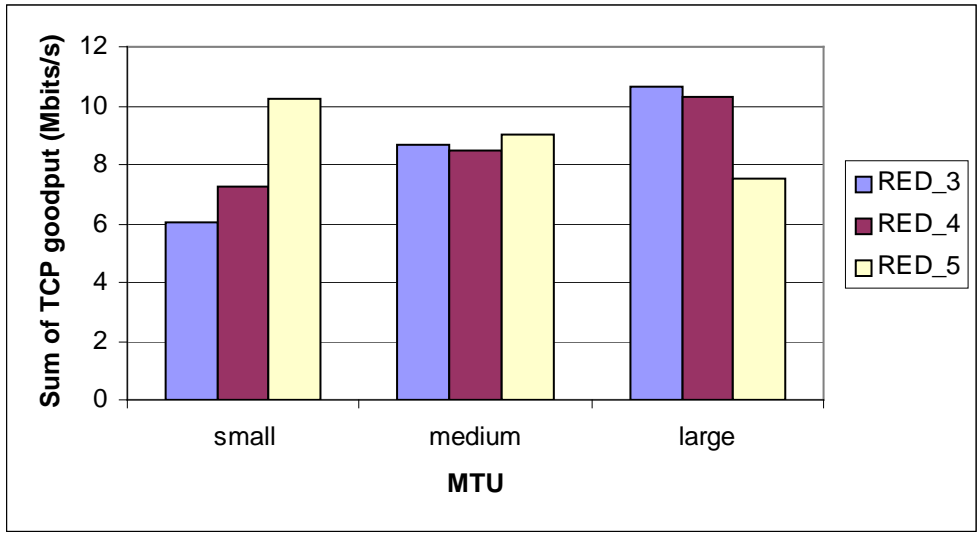

Figure 10: Obtained goodput for different MTUs, large propagation delay, TCP Reno

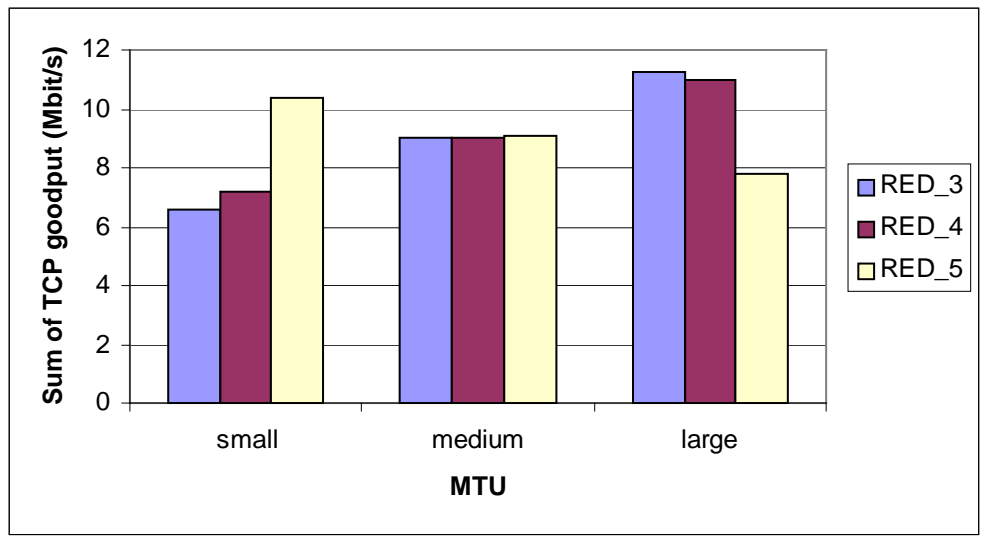

Figure 11: Obtained goodput for different MTUs, large propagation delay, TCP SACK

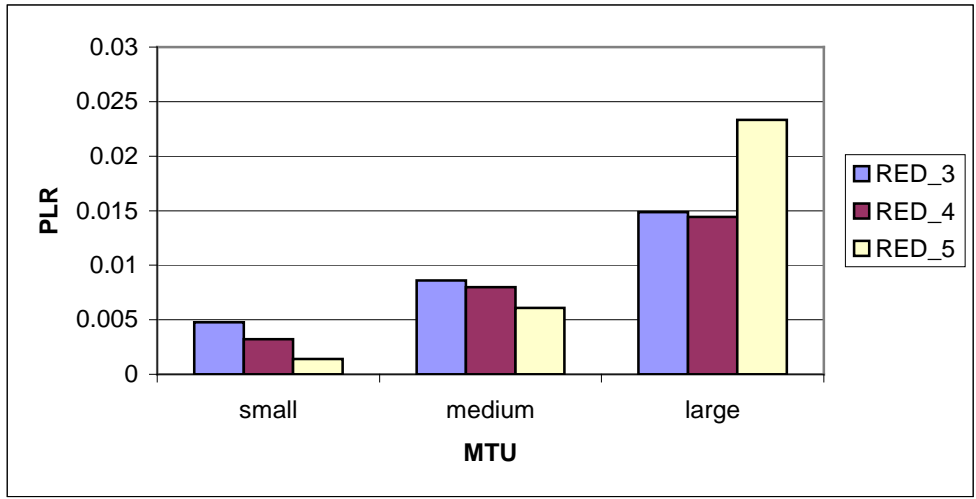

Figure 12: PLR for different MTUs, large propagation delay, TCP Reno

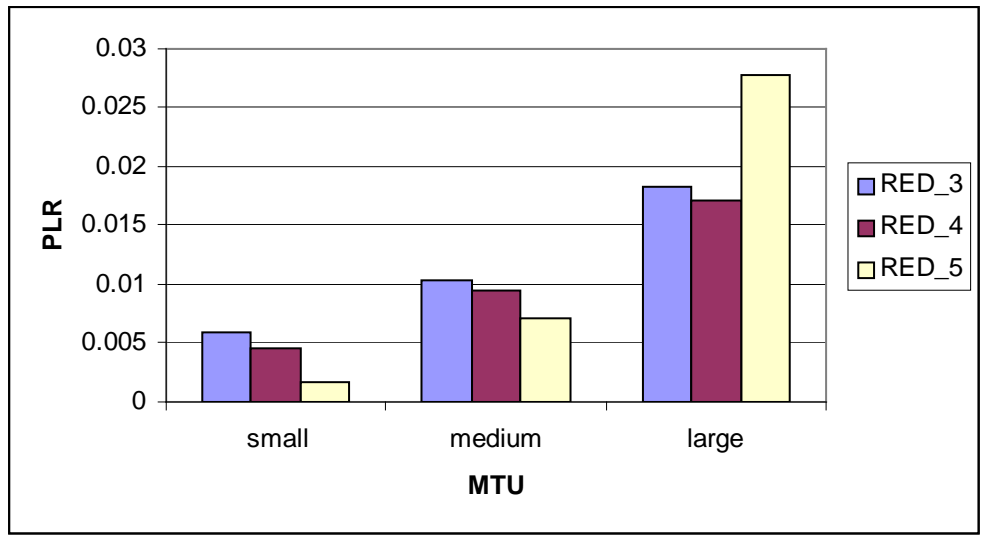

Figure 13: PLR for different MTUs, large propagation delay, TCP SACK

## 5/ Conclusions

The main results of our simulations can be summarized as follows:

- RED_1 can result in a too high a PLR for small Telnet packets,

- RED_2 could lead to a severe throughput collapse when the MTU is high,

- RED_3 gives good results in terms of loss differentiation and avoids low throughput (as it is the case with RED_2) for bulk transfers using large MTU values,

- RED_4 gives good results in terms of loss differentiation and fairness and in addition results in uniformly distributed drops,

- RED_5 is, from a theoretical point of view, the best RED variant to achieve fairness for TCP-friendly traffic.

Since the traffic in the Internet is a mixture of different packet sizes we strongly recommend the use of RED_4 or RED_5 which improve the PLR differentiation and do not result in throughput degradation for connections with large MTUs.

## 6/ References